\documentstyle[12pt,amssymb,amsmath]{article}
\newtheorem{theorem}{Theorem}[section]
\newtheorem{lemma}{Lemma}[section]

\begin{document}

\title{ {\bf Negaton and Positon  solutions of the soliton equation
with self-consistent sources }  }
\author{ {\bf  Yunbo Zeng \dag
    \hspace{1cm} Yijun Shao \dag \hspace{1cm} Weimin Xue \ddag} \\
    {\small {\it
  \dag  Department of Mathematical Sciences, Tsinghua University,
    Beijing 100084, China}}
    \\{\small{\it \ddag Mathematical Department, Hong Kong Baptist
    University, Kowloon, Hong Kong, China}} \\
    {\small {\it
    \dag E-mail: yzeng@math.tsinghua.edu.cn}}  }
\date{}
\maketitle
\renewcommand{\theequation}{\arabic{section}.\arabic{equation}}
%% \newcommand{\namelistlabel}[1]{\hfill\mbox{#1}}
%% \newenvironment{namelist}[1]{ %
%% \begin{list}{}
%%        {
%%          \let\makelabel\namelistlabel
%%          \settowidth{\labelwidth}{#1}
%%          \setlength{\leftmargin}{1.1\labelwidth}
%%        }
%%        }{ %
%%\end{list}}
%%%%%%%%%%%%%%%%%%macro%%%%%%%%%%%%%%%%%

\begin{abstract}
The KdV equation with self-consistent sources (KdVES) is used as a
model to illustrate the method. A generalized binary Darboux
transformation (GBDT) with an arbitrary time-dependent function
for the KdVES as well as the formula for $N$-times repeated GBDT
are presented. This GBDT provides non-auto-B\"{a}cklund
transformation between two KdV equations with different degrees of
sources and enable us to construct more general solutions with $N$
arbitrary $t$-dependent functions. By taking the special
$t$-function, we obtain multisoliton, multipositon, multinegaton,
multisoliton-positon, multinegaton-positon  and
multisoliton-negaton solutions of KdVES. Some properties of these
solutions are discussed.
\end{abstract}

\section{Introduction}
\setcounter{equation}{0}

 The soliton equations with
self-consistent sources (SESCS) have attracted some attension
(see, for example, \cite{mel89a}-\cite{mel88}). The SESCS can be
solved by the inverse scattering method and $N$-soliton solutions
of some SESCSs were obtained \cite{mel89a}-\cite{Lin2001}.
However, since the explicit time-part of the Lax representation
for SESCS was not found, the determination of the evolution for
scattering date was quite complicated in
\cite{mel89a}-\cite{mel88}. In recent years, we presented the
time-part of the Lax representation for SESCS by means of the
adjoint representation of soliton equation \cite{Zeng93,Zeng94}.
This enable us to determine the evolution of scattering date in a
simple and natural way \cite{Lin2001} and to construct the Darboux
transformation for SESCS \cite{Zeng2001,Zeng2002}. It was pointed
out in \cite{Zeng2001,Zeng2002} that the normal Darboux
transformation for SESCS which provides auto-B\"{a}cklund
transformation can not be used to construct solution of SESCS from
the trivial solution. In \cite{Zeng2001,Zeng2002} we presented
special kind of Binary Darboux transformation for some SESCSs
which offers non-auto-B\"{a}cklund transformation between soliton
equations with different degrees of sources and can be used to
obtained the $N$-soliton solutions. To our knowledge, no other
kinds of solution, except soliton solution, for SESCS are
investigated.

In recent years positon and negaton solution of soliton equations
have been wide studied (see \cite{Matveev2002} and references
therein). The positon solutions of soliton equation are long-range
analogues of solitons and slowly decreasing, oscillating
solutions, and possesses so-called supertransparent property: the
corresponding reflection coefficient is zero and the transmission
coefficient is unity \cite{Matveev2002}. The negaton solution of
KdV equation was studied in \cite{rasi96}.

 In this letter, we use the KdV equation with self-consistent sources (KdVES)
 as model to illustrate the idea. We present generalized binary
Darboux transformation (GBDT) with arbitrary $t$-dependent
functions for KdVES and the formula for $N$-times repeated GBDT
which contains $N$ arbitrary $t$-dependent functions. This GBDT
offers a non-auto-B\"{a}cklund transformation between KdV
equations with different degrees of sources and enables us to find
the more general solution with arbitrary  $t$-functions for KdVES.
By taking the special $t$-function, we obtain multisoliton,
multipositon, multinegaton, multisoliton-positon,
multisoliton-negaton and multipositon-negaton  solutions of KdVES.

This paper is organized as follows. In section 2, we derive the
GBDT with an arbitrary $t$-dependent function for the KdVES and
the formula for $N$-times repeated GBDT with $N$ arbitrary
$t$-dependent function. Using this GBDT gives rise to some kind of
general solutions of KdVES including multi-soliton solution as a
special case. In section 3 and 4, multi-positon and multi-negaton
solutions of KdVES are obtained, respectively. Finally, in section
5, multisoliton-positon, multisoliton-negaton and
multipositon-negaton  solutions of KdVES are presented.

\section{The generalized binary Darboux transformation}
 \setcounter{equation}{0}
 The KdV equation with
sources of degree $n$ (KdVES) is defined by
\cite{mel90a,Lati90,mel88,Lin2001,Zeng2001}
\begin{subequations}
\label{a1}
\begin{equation}
\label{a1a}
  u_t+6uu_x+u_{xxx}+4\sum_{j=1}^n\varphi_j\varphi_{j,x}=0,
\end{equation}
\begin{equation}
\label{a1b}
  \varphi_{j,xx}+(\lambda_j+u)\varphi_j=0,\quad j=1,\ldots,n,
\end{equation}
\end{subequations}
where $\lambda_j$ are distinct real constants. Let
$\Phi_n=(\varphi_1,\ldots,\varphi_n)$. The Lax representation for
(\ref{a1}) can be found from the adjoint representation for KdV
equation \cite{Zeng93,Zeng94,Zeng2001}
\begin{subequations}
\label{a2}
\begin{equation}
\label{a2a}
  \phi_{xx}+(\lambda+u)\phi=0,
\end{equation}
\begin{equation}
\label{a2b}
  \phi_t=A_n(\lambda,u,\Phi_n)\phi,
\end{equation}
\end{subequations}
where
$$A_n(\lambda,u,\Phi_n)\phi=u_x\phi+(4\lambda-2u)\phi_x
+\sum_{j=1}^n\frac{\varphi_j}{\lambda_j-\lambda}W(\varphi_j,\phi),$$
and $W(\varphi_j,\phi)\equiv\varphi_j\phi_x-\varphi_{j,x}\phi$ is
the usual Wronskian determinant.
 It is shown that the well-known Darboux transformation (DT) for
the KdV equation can be applied to the KdVES \cite{Zeng2001}. Let
$f$ be a solution of (\ref{a2}) with $\lambda=\xi$, then
(\ref{a2}) is covariant under the DT defined as \cite{Zeng2001}
\begin{subequations}
\label{a4}
\begin{equation}
\label{a4a}
\widetilde\phi=\frac{W(f,\phi)}{f},
\end{equation}
\begin{equation}
\label{a4b}
\widetilde u=u+2\partial_x^2\ln f,
\end{equation}
\begin{equation}
\label{a4c}
\widetilde\varphi_j=\frac{1}{\sqrt{\lambda_j-\xi}}\frac{W(f,\varphi_j)}{f},
\qquad j=1,...,n,
\end{equation}
\end{subequations}
i.e., $\widetilde\phi$, $\widetilde u$ and
$\widetilde\Phi_n=(\widetilde\varphi_1,\ldots,\widetilde\varphi_n)$ satisfy
\begin{subequations}
\label{a5}
\begin{equation}
\label{a5a}
  \widetilde\phi_{xx}+(\lambda+\widetilde u)\widetilde\phi=0,
\end{equation}
\begin{equation}
\label{a5b}
  \widetilde\phi_t=A_n(\lambda,\widetilde u,\widetilde\Phi_n)\widetilde\phi,
\end{equation}
\end{subequations}
and $\widetilde u$, $\widetilde\Phi_n$ is a new solution of
(\ref{a1}). Through this DT, we can find two linearly independent
solutions of (\ref{a5}) with $\lambda=\xi$. First, (\ref{a4a})
gives a solution of (\ref{a5}) with $\lambda=\xi$:
\begin{equation}
\label{a6}
  \widetilde f_1=\frac Cf,
\end{equation}
where $C$ is some constant.  Second, let $g$ be a solution of
(\ref{a2}) with $\lambda=\eta\neq\xi$, we define
$$ \omega(f,g)=\frac{W(f,g)}{\xi-\eta}.$$
According to (\ref{a4a}),
$$\widetilde g=\frac{1}{\xi-\eta}\frac{W(f,g)}{f}=\frac{1}{f}\omega(f,g)$$
 is a solution of (\ref{a5}) with $\lambda=\eta$.
For analytic $f=f(\xi)$ and let $g=f(\eta)$, we have
$$ \omega(f,f)\equiv\lim_{\eta\rightarrow\xi}\frac{W(f(\xi),f(\eta))}{\xi-\eta}
=-W(f,\partial_\xi f),$$ and
$$\widetilde f=\frac{1}{f}\omega(f,f)$$ is another solution of (\ref{a5})
with $\lambda=\xi$. Therefore
\begin{equation}
\label{a7} \widetilde h\equiv\widetilde f+\widetilde
f_1=\frac{1}{f}[C+\omega(f,f)]
\end{equation}
is also a solution of (\ref{a5}) with $\lambda=\xi$. Using $f$ and
$\widetilde h$ consecutively, the two-times action of DT
(\ref{a4}) yields the following binary DT
\begin{subequations}
\label{a8}
\begin{equation}
\label{a8a}
  \bar\phi=\frac{1}{\lambda-\xi}\frac{W(\widetilde h,\widetilde\phi)}{\widetilde h}
  =\phi-\frac{f}{C+\omega(f,f)}\omega(f,\phi),
\end{equation}
\begin{equation}
\label{a8b}
  \bar u=\widetilde u+2\partial_x^2\ln\widetilde h=u+2\partial_x^2\ln[C+\omega(f,f)],
\end{equation}
\begin{equation}
\label{a8c}
  \bar\varphi_j=\frac{1}{\sqrt{\lambda_j-\xi}}\frac{W(\widetilde h,\widetilde\varphi_j)}{\widetilde h}=\varphi_j-
  \frac{f}{C+\omega(f,f)}\omega(f,\varphi_j),\quad j=1,\ldots,n,
\end{equation}
\end{subequations}
then the system (\ref{a2}) is covariant under the binary DT
(\ref{a8}) and $\bar u$,
$\bar\Phi_n\equiv(\bar\varphi_1,\ldots,\bar\varphi_n)$ satisfies
the KdVES (\ref{a1}).

 Note that $\partial_x\omega(f,f)=f^2$,
$\partial_x\omega(f,\phi)=f\phi$ and
$\partial_x\omega(f,\varphi_j)=f\varphi_j$, the binary DT
(\ref{a8}) can be transformed into the original binary DT given in
\cite{Zeng2001}.

Substitution of (\ref{a8}) into (\ref{a2b}) gives
\begin{equation*}
\label{}
  \bar\phi_t=\left[\phi-\frac{f}{C+\omega(f,f)}\omega(f,\phi)\right]_t
\end{equation*}
\begin{equation}
\label{a10}
  =\phi_t-\frac{f_t\omega(f,\phi)}{C+\omega(f,f)}
  +\frac{f\partial_t\omega(f,f)}{[C+\omega(f,f)]^2}\omega(f,\phi)-\frac{f\partial_t\omega(f,\phi)}{C+\omega(f,f)}
  =A_n(\lambda,\bar u,\bar\Phi_n)\bar\phi.
\end{equation}
When substituting (\ref{a8}) into $A_n(\lambda,\bar
u,\bar\Phi_n)\bar\phi$, the last equality holds for any constant
$C$. In the expression of $A_n(\lambda,u,\Phi_n)\phi$, there is no
derivatives with respect to $t$. So the last equality holds when
$C$ is replaced by $e(t)$, an arbitrary $t$-function. We have the
following lemma.
\begin{lemma}
\label{lemma1} Given $u$, $\Phi_n$ a solution of (\ref{a1}), if
$f$ is a solution of (\ref{a2}) with $\lambda =\xi$, then the last
equality of (\ref{a10}) holds for $C=e(t)$
\end{lemma}

Obviously, under DT defined by (\ref{a8}) with $C$ replaced by
$e(t)$,  (\ref{a2a}) is still covariant, however, (\ref{a2b}) is
no longer covariant. In fact, we have
\begin{theorem}
\label{thm1} Given $u$, $\Phi_n$ as a solution of (\ref{a1}), let
$f$ be a solution of the system (\ref{a2}) with
$\lambda=\lambda_{n+1}$. Then, the generalized binary DT with an
arbitrary $t$-function defined by
\begin{subequations}
\label{a11}
\begin{equation}
\label{a11a}
\bar\phi=\phi-\frac{f}{e(t)+\omega(f,f)}\omega(f,\phi),
\end{equation}
\begin{equation}
\label{a11b}
\bar u=u+2\partial_x^2\ln[e(t)+\omega(f,f)],
\end{equation}
\begin{equation}
\label{a11c}
\bar\varphi_j=\varphi_j-\frac{f}{e(t)+\omega(f,f)}\omega(f,\varphi_j),\quad j=1,\ldots,n,
\end{equation}
and
\begin{equation}
\label{a11d}
\bar\varphi_{n+1}=\frac{\sqrt{e'(t)}f}{e(t)+\omega(f,f)}
\end{equation}
\end{subequations}
transforms (\ref{a2}) into
\begin{subequations}
\label{a12}
\begin{equation}
\label{a12a}
\bar\phi_{xx}+(\lambda+\bar u)\bar\phi=0,
\end{equation}
\begin{equation}
\label{a12b}
  \bar\phi_t=A_{n+1}(\lambda,\bar
u,\bar\Phi_{n+1})\bar\phi,
\end{equation}
\end{subequations}
and $\bar u$,
$\bar\Phi_{n+1}\equiv(\bar\varphi_1,\ldots,\bar\varphi_{n+1})$,
satisfy the KdV equation with sources of degree $n+1$
\begin{subequations}
\label{a13}
\begin{equation}
\label{a13a}
  \bar u_t+6\bar u\bar u_x+\bar u_{xxx}+4\sum_{j=1}^{n+1}\bar\varphi_j\bar\varphi_{j,x}=0,
\end{equation}
\begin{equation}
\label{a13b}
  \bar\varphi_{j,xx}+(\lambda_j+\bar u)\bar\varphi_j=0,\quad j=1,\ldots,n+1,
\end{equation}
\end{subequations}
\end{theorem}
\textbf{Proof}.  $\widetilde h$ defined by (\ref{a7}) with $C$
replaced by $e(t)$  still satisfies (\ref{a5a}). This implies that
(\ref{a12a}) and (\ref{a13b}) hold. Substituting (\ref{a11a}) into
the left-hand side of (\ref{a12b}) and using  the Lemma 2.1 gives
rise to
$$\bar\phi_t=\left[\phi-\frac{f}{e(t)+\omega(f,f)}\omega(f,\phi)\right]_t
=\phi_t-\frac{f_t\omega(f,\phi)}{e(t)+\omega(f,f)}
  +\frac{f[e'(t)+\partial_t\omega(f,f)]}{[e(t)+\omega(f,f)]^2}\omega(f,\phi)
$$
$$
  -\frac{f\partial_t\omega(f,\phi)}{e(t)+\omega(f,f)}
  =A_n(\lambda,\bar u,\bar\Phi_n)\bar\phi
  +\frac{e'(t)\omega(f,\phi)f}{[e(t)+\omega(f,f)]^2}
=A_{n+1}(\lambda,\bar u,\bar\Phi_{n+1})\bar\phi,$$ where we have
use the formula
$$
  \frac{W(\bar\varphi_{n+1},\bar\phi)}{\lambda_{n+1}-\lambda}
  =\frac{\sqrt{e'(t)}\omega(f,\phi)}{e(t)+\omega(f,f)}.
$$
Then the compatibility condition of (\ref{a12}) leads to
(\ref{a13a}). This completes the proof.

The generalized binary DT (GBDT) defined by (\ref{a11}) contains
an arbitrary $t$-function. The flexibility of the choices of
$e(t)$ and $f$ enables us to construct some kind of general
solutions with arbitrary $t$-functions of the KdVES some of that
can not be constructed through the original binary DT.

 For $m$
solutions of (\ref{a2}), $g_1,\ldots,g_m$ and $m$ arbitrary
$t$-functions $e_1(t),\ldots,e_m(t)$ , we define two types of
Wronskian determinant
$$W_1(g_1,\ldots,g_m;e_1,\ldots,e_m)=\det F, \quad W_2(g_1,\ldots,g_m;e_1,\ldots,e_{m-1})=\det
G$$ where
$$ F_{ij}=\delta_{ij}e_i(t)+\omega(g_i,g_j),\quad i,j=1,\ldots,m,$$
$$ G_{ij}=\delta_{ij}e_i(t)+\omega(g_i,g_j),\quad i=1,\ldots,m-1,\quad j=1,\ldots,m,$$
$$ G_{mj}=g_j,\quad j=1,\ldots,m.$$
We have the following formula of $N$-times repeated GBDT.
\begin{theorem}
\label{thm2} Given $u$, $\Phi_n$ as a solution of (\ref{a1}), let
$f_1,\ldots,f_N$ be solutions of (\ref{a2}) with
$\lambda=\lambda_{n+1},\ldots,\lambda_{n+N}$, respectively. Then
the $N$-times repeated GBDT with $N$-arbitrary $t$-functions
$e_1(t),\ldots,e_N(t)$ defined by
\begin{subequations}
\label{a14}
\begin{equation}
\label{a14a}
\bar\phi=\frac{W_2(f_1,\ldots,f_N,\phi;e_1,\ldots,e_N)}{W_1(f_1,\ldots,f_N;e_1,\ldots,e_N)},
\end{equation}
\begin{equation}
\label{a14b}
\bar u=u+2\partial_x^2\ln W_1(f_1,\ldots,f_N;e_1,\ldots,e_N),
\end{equation}
\begin{equation}
\label{a14c}
\bar\varphi_j=\frac{W_2(f_1,\ldots,f_N,\varphi_j;e_1,\ldots,e_N)}{W_1(f_1,\ldots,f_N;e_1,\ldots,e_N)},\quad
j=1,\ldots,n,
\end{equation}
and
\begin{equation}
\label{a14d}
\bar\varphi_{n+j}=\frac{\sqrt{e'_j(t)}W_2(f_1,\ldots,f_{j-1},f_{j+1},
\ldots,f_N,f_j;e_1,\ldots,e_{j-1},e_{j+1},\ldots,e_N)}{W_1(f_1,\ldots,f_N;e_1,
\ldots,e_N)},\quad
j=1,\ldots,N
\end{equation}
\end{subequations}
transforms (\ref{a2}) into (\ref{a2}) with $n$ replaced by $n+N$
and $\bar u$, $\overline\Phi_{n+N}$ satisfy the KdVES of degree
$n+N$, i.e. (\ref{a1}) with $n$ replaced by $n+N$.
\end{theorem}
The proof of this theorem is completely similar to that given in
\cite{Zeng2001} and we omit it.

\textbf{Example: $N$-soliton solution.}

 We take $u=0$ as the initial solution of
(\ref{a1}) with $n=0$ and let $\lambda_j=-\kappa_j^2<0,
\kappa_j>0$, $j=1,\ldots,N,$
$$ f_j=e^{\kappa_jx-4\kappa_j^3t},\quad e_j(t)=e^{2\alpha_jt},
\quad j=1,\ldots,N,$$ then
$$\omega(f_i,f_j)=\frac 1{\kappa_i+\kappa_j}e^{(\kappa_i+\kappa_j)x
-4(\kappa_i^3+\kappa_j^3)t},\quad i,j=1,\ldots,N,$$
 the $N$-soliton solutions of
(\ref{a1}) with $n=N$ and $\lambda_j=-\kappa_j^2<0,$,
$j=1,\ldots,N,$ is given by
\begin{subequations}
\label{a15}
\begin{equation}
\label{a15a}
  u=2\partial_x^2\ln W_1(f_1,\ldots,f_N;e_1,\ldots,e_N),
\end{equation}
\begin{equation}
\label{a15b}
  \varphi_j=\frac{\sqrt{a_j}W_2(f_1,\ldots,f_{j-1},f_{j+1},
  \ldots,f_N,f_j;e_1,\ldots,e_{j-1},e_{j+1},\ldots,e_N)}{W_1(f_1,\ldots,f_N;e_1,\ldots,e_N)},\quad
  j=1,\ldots,N,
\end{equation}
\end{subequations}
 which was obtained in
\cite{mel88,Lin2001,Zeng2001}.

\section{Positon solutions}
\setcounter{equation}{0}
  Hereafter we
always take simple and special choice of $e(t)$ as
\begin{equation}
\label{b3} e_j(t)=a_jt+b_j,
\end{equation}
where $a_j\neq 0$ and $b_j$ are real constants.
\textbf{3.1
One-positon solution and the supertranseparency.}
 We take $u=0$ as
the initial solution of (\ref{a1}) with $n=0$. Let $f$ be an
oscillating solution of (\ref{a2}) with $u=0$, $n=0$ and
$\lambda=\lambda_1=\kappa^2>0, \kappa>0$,
\begin{equation}
\label{b2}
  f=\sin\Theta, \quad \Theta=\kappa(x+x_1+4\kappa^2t),
\end{equation}
where $x_1=x_1(\kappa)$ is an real differential function of
$\kappa$. Then the GBDT (\ref{a11}) gives
\begin{subequations}
\label{b1}
\begin{equation}
\label{b1a}
  u=2\partial_x^2\ln(2\kappa\gamma-\sin2\Theta)=
  \frac{32\kappa^2\sin\Theta(\kappa\gamma\cos\Theta-\sin\Theta)}
  {(2\kappa\gamma-\sin2\Theta)^2},
\end{equation}
\begin{equation}
\label{b1b}
  \varphi_1=
  \frac{4\kappa\sqrt{a}\sin\Theta}{2\kappa\gamma-\sin2\Theta},
\end{equation}
\end{subequations}
with
$$\gamma=\partial_\kappa\Theta+2e(t)=
x+\tilde x_1+(12\kappa^2+2a)t+2b,\quad \tilde
x_1=x_1+\kappa\partial_\kappa x_1(\kappa),$$
 which gives the one-positon solution of KdVES (\ref{a1}) with
$n=1, \lambda_1=\kappa^2$  corresponding to the one-positon
solution for the KdV equation in \cite{Matveev2002,rasi96}.

 The solution of linear system (\ref{a2}) with $n=1, \lambda=k^2, \lambda_1=\kappa^2$
 $u$ and $\varphi_1$ given by (\ref{b1}) is
\begin{equation}
\label{b5}
  \phi(x,k)
  =\left(-k^2+\frac{4ik\kappa\sin^2\Theta}{\sin2\Theta-2\kappa\gamma}
  -\kappa^2\frac{\sin2\Theta+2\kappa\gamma}{\sin2\Theta-2\kappa\gamma}\right)
  \frac{e^{ikx+4ik^3t}}{-k^2+\kappa^2}.
\end{equation}

Based on formulas (\ref{b1}) and (\ref{b5}), we can analyze the
basic features of the one-positon solution of
 ({\ref{a1}) in the same way as in \cite{Matveev2002}. We can
conclude that the one-positon solution of (\ref{a1}) with $n=1$
has the same shape, the same asymptotic behavior when
$x\rightarrow\pm\infty$ and the same scattering data as the one
positon solution of the KdV equation: it is long-range analogues
of solitons of the KdVES and is slowly decreasing, oscillating
solutions. Similarly under a proper choice of the scattering data,
the corresponding reflection coefficient is zero and the
transmission coefficient is unity.

 \textbf{3.2 Two-positon
solution and multi-positon  solutions.}

The two-positon solution of (\ref{a1}) with $n=2,
\lambda_j=\kappa_j^2>0, \kappa_j>0, j=1,2,$ is given by
(\ref{a15}) with $N=2, e_j=a_jt+b_j$
$$ f_j=\sin\Theta_j,\quad \Theta_j=\kappa_j(x+x_j+4\kappa_j^2t), \quad j=1,2, $$
$$  W_1(f_1,f_2;e_1,e_2)=(16\kappa_1\kappa_2)^{-1}(2\kappa_1\gamma_1-\sin2\Theta_1)
  (2\kappa_2\gamma_2-\sin2\Theta_2)$$
\begin{subequations}
\label{b7}
\begin{equation}
\label{b7a}
  -(\kappa_1^2-\kappa_2^2)^{-2}(\kappa_2\sin\Theta_1\cos\Theta_2-\kappa_1\sin\Theta_2\cos\Theta_1)^2,
\end{equation}
$$W_2(f_2,f_1;e_2)=(4\kappa_2)^{-1}\sin\Theta_1(2\kappa_2\gamma_2-\sin2\Theta_2)$$
\begin{equation}
\label{b7b}
  -(\kappa_1^2-\kappa_2^2)^{-1}\sin\Theta_2(\kappa_2\sin\Theta_1\cos\Theta_2-\kappa_1\sin\Theta_2\cos\Theta_1),
\end{equation}
$$W_2(f_1,f_2;e_1)=(4\kappa_1)^{-1}\sin\Theta_2(2\kappa_1\gamma_1-\sin2\Theta_1)$$
\begin{equation}
\label{b7c}
  -(\kappa_1^2-\kappa_2^2)^{-1}\sin\Theta_1(\kappa_2\sin\Theta_1\cos\Theta_2-\kappa_1\sin\Theta_2\cos\Theta_1),
\end{equation}
\end{subequations}
$$ \gamma_j=x+\widetilde x_j+(12\kappa_j^2+2a_j)t+2b_j,\quad \widetilde
x_j=x_j+\kappa_j\partial_{\kappa_j} x_j(\kappa_j),\quad j=1,2.$$
 Using
(\ref{b7}), we obtain the asymptotic behavior of the solution for
fixed $\gamma_1$ as $t\rightarrow\pm\infty$ (which implies
$\gamma_2\rightarrow\infty$)
$$
  u=2\partial_x^2\ln(2\kappa_1\gamma_1-\sin2\Theta_1)[1+O(\gamma_2^{-1})],
$$
$$
  \varphi_1=\frac{4\kappa_1\sqrt{a_1}\sin\Theta_1}{2\kappa_1\gamma_1
  -\sin2\Theta_1}[1+O(\gamma_2^{-1})],
  \quad \varphi_2=O(\gamma_2^{-1}).
$$
In the asymptotic domain where $\gamma_2$ is fixed and
$t\rightarrow\pm\infty$($\gamma_1\rightarrow\infty$), we have
$$
  u=2\partial_x^2\ln(2\kappa_2\gamma_2-\sin2\Theta_2)[1+O(\gamma_1^{-1})],
$$
$$
  \varphi_1=O(\gamma_1^{-1}),\quad \varphi_2=\frac{4\kappa_2\sqrt{a_2}
  \sin\Theta_2}{2\kappa_2\gamma_2-\sin2\Theta_2}[1+O(\gamma_1^{-1})].
$$
Thus we have proved that the two positons are totally insensitive
to the mutual collision, even without additional phase shifts
which is intrinsic for the collision of two solitons. Calculating
the corresponding solution of system (\ref{a2}), we can prove that
potential is also supertransparent.

The $N$-positon  solution of (\ref{a1}) with $n=N,
\lambda_j=\kappa_j^2>0, \kappa_j>0, j=1,\ldots,N,$ is given by
(\ref{a15}) with $ e_j=a_jt+b_j,$
$$ f_j=\sin\Theta_j,\quad \Theta_j=\kappa_j(x+x_j+4\kappa_j^2t), \quad j=1\ldots,N, $$
$$\omega(f_i,f_j)=(\kappa_i^2-\kappa_j^2)^{-2}(\kappa_j\sin\Theta_i\cos\Theta_j
-\kappa_i\sin\Theta_j\cos\Theta_i)^2.$$

Analogously, we will see that the $N$-positon solution at large
time decays into the sum of $N$ free positons and it is also
supertransparent.

\section{Negaton  solutions}
\setcounter{equation}{0}
 \textbf{4.1 One-negaton solution.}
 Let $\lambda_1=-\kappa^2<0, \kappa>0$ and $f$ be a solution of
(\ref{a2}) with $u=0$, $n=0$ and $\lambda=\lambda_1$,
\begin{equation}
\label{c2}
  f=\sinh\Theta, \quad \Theta=\kappa(x+x_1-4\kappa^2t).
\end{equation}
Then the GBDT (\ref{a11b}) and (\ref{a11d}) with $ e(t)=at+b$
gives
\begin{subequations}
\label{c1}
\begin{equation}
\label{c1a}
u=2\partial_x^2\ln(\kappa\gamma-\sinh\Theta\cosh\Theta)=
  \frac{8\kappa^2\sinh\Theta(\sinh\Theta-\kappa\gamma\cosh\Theta)}
  {(\kappa\gamma-\sinh\Theta\cosh\Theta)^2},
\end{equation}
\begin{equation}
\label{c1b}
  \varphi_1=
 \frac{2\kappa\sqrt{a}\sinh\Theta}{\kappa\gamma-\sinh\Theta\cosh\Theta},
\end{equation}
where
\begin{equation}
\label{c1c} \gamma=x+x_1+\kappa\partial_\kappa x_1
-(12\kappa^2-2a)t+2b.
\end{equation}
\end{subequations}
 (\ref{c1}) gives the [S] one-negaton solution of (\ref{a1}) with $n=1$ and
$\lambda_1=-\kappa^2<0$ which  corresponds to the [S] one-negaton
solution for the KdV equation in \cite{rasi96}.

When $t$ is fixed, then we have
$$  u\sim 8\kappa^2(\frac 1{\cosh^2\Theta}-\frac{\kappa \gamma}{\sinh\Theta\cosh\Theta})
\rightarrow 0,\quad\quad
 \varphi_1\sim -\frac{\kappa\sqrt{a}}{\cosh\Theta}\rightarrow 0,
 \quad\quad x\rightarrow\pm\infty.$$
For fixed $x$, we have the same formula when
$t\rightarrow\pm\infty.$

As a function of $x$, $u$ has a second-order pole and $\varphi_1$
has a first-order pole which locate at the same point $x=x_p(t)$
determined by the equation
$\sinh\Theta\cosh\Theta-\kappa\gamma=0$. Also it is easy to that
$u(x,t)$ has two zeros and $\varphi_1(x,t)$ has one zero. The
shape and the motion of $u(x,t)$ is the same as that described in
\cite{rasi96}.

Similarly, if we take $f=\cosh\Theta$, we can obtain the [C]
one-negaton.

\textbf{4.2 Two-negaton solution and multi-negaton solutions.}

The [S] two-negaton solution of (\ref{a1}) with $n=2$ and
$\lambda_j=-\kappa_j^2<0, \kappa_j>0, j=1,2$ is given by
(\ref{a15}) with $N=2$
$$ f_j=\sinh\Theta_j,\quad \Theta_j=\kappa_j(x+x_j-4\kappa_j^2t), \quad
e_j(t)=a_jt+b_j, \quad j=1,2,$$
$$
  W_1(f_1,f_2;e_1,e_2)=(4\kappa_1\kappa_2)^{-1}(\kappa_1\gamma_1-\sinh\Theta_1\cosh\Theta_1)
  (\kappa_2\gamma_2-\sinh\Theta_2\cosh\Theta_2),
$$
\begin{equation}
\label{c7}
  -(\kappa_1^2-\kappa_2^2)^{-2}(\kappa_2\sinh\Theta_1\cosh\Theta_2-\kappa_1\sinh\Theta_2
  \cosh\Theta_1)^2,
\end{equation}
$$W_2(f_2,f_1;e_2)=(2\kappa_2)^{-1}\sinh\Theta_1(\kappa_2\gamma_2-\sinh\Theta_2\cosh\Theta_2)$$
\begin{equation}
\label{c8}
  +(\kappa_1^2-\kappa_2^2)^{-1}\sinh\Theta_2(\kappa_2\sinh\Theta_1\cosh\Theta_2-\kappa_1
  \sinh\Theta_2\cosh\Theta_1),
\end{equation}
$$W_2(f_1,f_2;e_1)=(2\kappa_1)^{-1}\sinh\Theta_2(\kappa_1\gamma_1-\sinh\Theta_1\cosh\Theta_1)$$
\begin{equation}
\label{c9}
+(\kappa_1^2-\kappa_2^2)^{-1}\sinh\Theta_1(\kappa_2\sinh\Theta_1\cosh\Theta_2-\kappa_1
\sinh\Theta_2\cosh\Theta_1),
\end{equation}
$$ \gamma_j=x+\widetilde x_j-12\kappa_j^2t+2a_jt+b_j,\quad \widetilde
x_j=x_j+\kappa_j\partial_{\kappa_j} x_j(\kappa_j), \quad Im
x_j=0, \quad j=1,2.$$

In the domain where $x+x_1-4\kappa_1^2t$ is fixed and
$t\rightarrow \pm\infty$, the asymptotic solution is
$$
  u=2\partial_x^2\ln(\kappa_1\gamma_1-\sinh\Theta_1\cosh\Theta_1)[1+O(t^{-1})],
$$
$$
  \varphi_1=\frac{2\kappa_1\sqrt{a_1}\sinh\Theta_1}{\kappa_1\gamma_1-\sinh\Theta_1\cosh\Theta_1}
  [1+O(t^{-1})],
  \quad \varphi_2=O(t^{-1}).
$$

In the asymptotic domain where $x+x_2-4\kappa_2^2t$ is fixed and
$t\rightarrow\pm\infty$, we have
$$
  u=2\partial_x^2\ln(\kappa_2\gamma_2-\sinh\Theta_2\cosh\Theta_2)[1+O(t^{-1})],
$$
$$
  \varphi_1=O(t^{-1}),\quad \varphi_2=\frac{2\kappa_2\sqrt{a_2}\sinh\Theta_2}
  {\kappa_2\gamma_2-\sinh\Theta_2\cosh\Theta_2}[1+O(t^{-1})].
$$
This estimates show that in the indicated domain, the leading
terms of the asymptotic [S] two-negaton solution is a standard [S]
one-negaton solution. In other words, negatons are totally
insensitive to the mutual collision, even without additional phase
shifts in contrast to  the solitons collision case.

Similarly we can construct the [C] two-negaton and [SC]
two-negaton and find the same property.

 The [S] $N$-negaton
solution of (\ref{a1}) with $n=N$ and $\lambda_j=-\kappa_j^2<0,
\kappa_j>0$, $j=1,\ldots,N$ is given by (\ref{a15}) with
$e_j=a_jt+b_j$
$$ f_j=\sinh\Theta_j,\quad \Theta_j=\kappa_j(x+x_j-4\kappa_j^2t),\quad
{\rm{Im}}x_j=0, \quad j=1\ldots,N. $$

Analogously, we will see that the [S] $N$-negaton solution at
large time decays into the sum of $N$ [S] free negatons.

\section{Multisoliton-positon,  Multisoliton-negaton and\\
Multipositon-negaton solutions} \setcounter{equation}{0}

Like the KdV equation, the  KdVES also has Multisoliton-positon,
Multisoliton-negaton and Multipositon-negaton  solutions. The
$N$-positon $M$-soliton  solutions of (\ref{a1}) with $n=N+M$ and
$\lambda_j=\kappa_j^2>0$, $j=1,\ldots,N$,
$\lambda_{N+j}=-\kappa_{N+j}^2<0$, $j=1,\ldots,M$ are given by
(\ref{a15}) with $N$ replaced by $N+M$ and
$$ f_j=\sin\Theta_j,\quad \Theta_j=\kappa_j(x+x_j+4\kappa_j^2t),\quad
\kappa_j>0, \quad j=1\ldots,N, $$
$$ f_{N+j}=e^{\kappa_{N+j}(x-4\kappa_{N+j}^2t)},\quad
\kappa_{N+j}>0, \quad j=1\ldots,M.$$

The $N$-negaton $M$-soliton  solution of (\ref{a1}) with $n=N+M$
and $\lambda_j=-\kappa_j^2>0$, $j=1,\ldots,N+M$,  is given by
(\ref{a15}) with $N$ replaced by $N+M$ and
$$ f_j=\sinh(\cosh)\Theta_j,\quad \Theta_j=\kappa_j(x+x_j-4\kappa_j^2t),\quad
\kappa_j>0, \quad j=1\ldots,N, $$
$$ f_{N+j}=e^{\kappa_{N+j}(x-4\kappa_{N+j}^2t)},\quad
\kappa_{N+j}>0, \quad j=1\ldots,M.$$

The $N$-positon $M$-negaton  solution of (\ref{a1}) with $n=N+M$
and $\lambda_j=\kappa_j^2>0$, $j=1,\ldots,N$,
$\lambda_{N+j}=-\kappa_{N+j}^2<0$, $j=1,\ldots,M$  is given by
(\ref{a15}) with $N$ replaced by $N+M$ and
$$ f_j=\sin\Theta_j,\quad \Theta_j=\kappa_j(x+x_j+4\kappa_j^2t),\quad
\kappa_j>0, \quad j=1\ldots,N, $$
$$ f_{N+j}=\sinh (\cosh)\Theta_{N+j},\quad \Theta_{N+j}=\kappa_{N+j}(x+x_{N+j}
-4\kappa_{N+j}^2t),\quad \kappa_{N+j}>0, \quad j=1\ldots,M.$$

 We can analyze the interaction of the soliton and the positon, the soliton
 and the negaton,
 the positon and the negaton in
a similar way as in \cite{Matveev2002}. We would like to point out
that the results of the analysis will be almost the same as in
\cite{Matveev2002} and  we omit it.

\section{Conclusions}
\setcounter{equation}{0}
 We present $N$-times repeated GBDT with $N$ arbitrary
 $t$-functions which provides no-auto-B\"{a}cklund transformation
 between two KdV equations with $n$-degrees of sources and
 $n+N$-degrees of sources. This $N$-times repeated GBDT enables us
 to construct some kind of general solutions with $N$ arbitrary
 $t$-functions for KdVES. By taking special choice of the
 $t$-functions we obtain the Multi-soliton,  Multi-negaton
Multi-positon, Multisoliton-positon,  Multisoliton-negaton and
Multipositon-negaton solutions for KdVES. The method can be
applied to other SESCSs.

\hskip\parindent

\section*{ Acknowledgment }
This work is supported by the Special Funds for Chinese Major
Basic Research Project "Nonlinear Science" and Research Grants
Council of Hong Kong (HKBU/2047/02P).

\hskip\parindent

\end{document}